\documentclass[epj]{webofc}
\usepackage[utf8]{inputenc}
\usepackage[varg]{txfonts}   % Web of Conferences font
\usepackage{booktabs}
\usepackage{xcolor}
\definecolor{darkred}{rgb}{0.4,0.0,0.0}
\definecolor{darkgreen}{rgb}{0.0,0.4,0.0}
\definecolor{darkblue}{rgb}{0.0,0.0,0.4}
\usepackage[bookmarks,linktocpage,colorlinks,
    linkcolor = darkred,
    urlcolor  = darkblue,
    citecolor = darkgreen]{hyperref}
\wocname{EPJ Web of Conferences}
\woctitle{Lattice2017}
\newcommand{\be}{\begin{equation}}
\newcommand{\ee}{\end{equation}}
\newcommand{\ep}{\epsilon}

\begin{document}
%%%%%%%%%%%%%%%%%%%%%%%%%%%%%%%%%%%%%%%%%%%%%%%%%%%%%%%%%%%%%%%%%%%%%%%%%%%%%
%
\selectlanguage{english}
%----------------------------------------------------------------------------
\title{%
A new method for the beta function in the chiral symmetry broken phase
}
%----------------------------------------------------------------------------
\author{%
\firstname{Zoltan} \lastname{Fodor}\inst{1,2} \and
\firstname{Kieran} \lastname{Holland}\inst{3}\fnsep\thanks{Speaker} \and
\firstname{Julius}  \lastname{Kuti}\inst{4} \and
\firstname{Daniel}  \lastname{Nogradi}\inst{5,6} \and
\firstname{Chik Him}  \lastname{Wong}\inst{1}
% etc.
}
%----------------------------------------------------------------------------
\institute{%
University of Wuppertal, Department of Physics, Wuppertal D-42097, Germany
\and
Juelich Supercomputing Center, Forschungszentrum Juelich, Juelich D-52425, Germany
\and
University of the Pacific, 3601 Pacific Ave, Stockton CA 95211, USA
\and
University of California, San Diego, 9500 Gilman Drive, La Jolla CA 92093, USA
\and
E\"{o}tv\"{o}s  University,  Institute  for  Theoretical  Physics,
MTA-ELTE Lendulet Lattice Gauge Theory Research Group, Budapest 1117, Hungary
\and 
Universidad Autonoma, IFT UAM/CSIC and Departamento de Fisica Teorica,
28049 Madrid, Spain
}
%----------------------------------------------------------------------------
\abstract{%
  We describe a new method to determine non-perturbatively the beta function of a gauge theory 
  using lattice simulations in the p-regime of the theory. This complements alternative  
  measurements of the beta function working directly at zero fermion mass and bridges the gap 
  between the weak coupling perturbative regime and the strong coupling regime relevant to the
  mass spectrum of the theory. We apply this method to ${\mathrm {SU(3)} }$ gauge theory with two
  fermion flavors in the 2-index symmetric (sextet) representation. We find that the beta function is
  small but non-zero at the renormalized coupling value $g^2 = 6.7$, consistent with our previous
  independent investigation using simulations directly at zero fermion mass. The model continues to be
  a very interesting explicit realization of the near-conformal composite Higgs paradigm which could
  be relevant for Beyond Standard Model phenomenology. 
}
%----------------------------------------------------------------------------
\maketitle
%----------------------------------------------------------------------------
\section{Introduction}\label{intro}

In the search for near-conformal composite Higgs theories, non-perturbative lattice determinations of the beta function of the renormalized gauge coupling have played a crucial role. The recent development of the gradient flow~\cite{Luscher:2010iy,Narayanan:2006rf,Lohmayer:2011si,Luscher:2011bx} has added a new level of precision allowing for very accurate measurement of renormalized quantities. However lattice simulations of Beyond Standard Model (BSM) theories with an increased number of fermion flavors, like $N_f = 12$, or fermion representations other than the fundamental can lead to a significant increase in computational effort compared to simulations of QCD. 

To determine if a given model is infrared conformal or not, one has to know the behavior in the chiral limit. For beta function studies, that typically leads to working directly at zero fermion mass, with a particular choice of boundary conditions. There are also complementary studies of the particle spectrum of such theories, where the fermion mass is varied to see if e.g.~chiral symmetry appears to be spontaneously broken in the massless limit generating a set of Goldstone bosons, or if a light composite scalar particle, perhaps a Higgs impostor, exists in such a model. Given the large computational resources each such study requires, a beta function measurement which can take advantage of pre-exisiting particle spectrum type gauge ensembles would be very valuable, since {\bf (a)} it would involve negligible additional computational cost, {\bf (b)} the beta function would be measured at renormalized gauge couplings strong enough to see if chiral symmetry could be spontaneously broken in the chiral limit, and {\bf (c)} it would complement independent beta function measurements from simulations directly at zero fermion mass. In this report we describe such a technique. We apply it in the context of near-conformal gauge theories, the method can just as well be applied to other gauge theories such as QCD.

\section{Gradient flow and step-scaling in finite volume}\label{gradient}

The gradient flow $d A_\mu/dt = D_\nu F_{\nu \mu}$ defines the gauge field $A_\mu(t)$ at flow time $t$. Perturbatively, the action density $E = (F_{\mu \nu}^a)^2/4$ has an expectation value
\be
\langle E \rangle = \frac{3(N^2-1) g^2}{128 \pi^2 t^2} \left\{ 1 + \overline{c_1} g^2 + {\cal O}(g^4) \right\}
\label{eq1}
\ee
in the $\overline{\mathrm {MS}}$ scheme for ${\mathrm {SU(N)}}$ gauge theory where the renormalized coupling $g$ is defined at the renormalization group scale $\mu = 1/\sqrt{8t}$. This motivates a non-perturbative definition of the renormalized coupling
\be
g^2(t) \equiv \frac{1}{\cal N} \left( \frac{128 \pi^2}{3(N^2-1)} \right) t^2 \langle E \rangle_{\mathrm {latt}},
\label{eq2}
\ee
where the expectation value of the action density at flow time $t$ is measured via lattice simulations and the normalization factor ${\cal {N}}$ depends on the choice of boundary conditions. As the action density is a bulk quantity, the observable $\langle E \rangle$ can be measured non-perturbatively very precisely.

One way to measure the beta function in finite volume is via step-scaling: in a physical volume $L^4$, the flow is adjusted holding $c = \sqrt{8t}/L$ fixed, each choice of $c$ corresponding to a particular renormalization group (RG) scheme. The RG scale $\mu$ is now in terms of the only remaining scale $L$. For a given lattice volume $(L/a)^4$ the bare gauge coupling (and hence the lattice spacing) is adjusted such that the renormalized coupling has a chosen fixed value e.g.~$g^2_c(L/a)=6$. Keeping the lattice spacing $a$ fixed, a second simulation on a larger volume e.g.~$(sL/a)^4$ with $s=2$ gives the discrete step $\beta(g^2_c) = \{g^2_c(sL/a) - g^2_c(L/a)\}/\log(s^2)$ i.e.~the response of the gauge coupling as the RG scale is changed by a finite amount. In this context {\it discrete} has nothing to do with the lattice discretization. However the beta function will contain lattice artifacts which must be removed. To take the continuum limit, the procedure is repeated for a sequence of lattice volumes e.g.~$L/a = 16, 18, 20, 24, 28$ on each of which $g^2_c(L/a) = 6$ is tuned via the bare coupling and larger volumes e.g.~$2L/a = 32, 36, 40, 48, 56$ from which the discrete step is measured and the limit $a/L \rightarrow 0$ is obtained. The final result is the continuum finite-step beta function in finite volume. This approach, widely used in QCD, has already been applied in the context of near-conformal gauge theories~\cite{Fodor:2012td,Hasenfratz:2014rna,Fodor:2015baa,Lin:2015zpa,Appelquist:2009ty,Hietanen:2009az,Hayakawa:2013yfa}.

\section{Beta function in infinite volume}\label{beta}

The main message of this report is to describe an alternative approach. Since the gradient flow defines a renormalized coupling $g^2(t)$ at any flow time $t$, one can also directly measure on the same ensemble of gauge configurations the derivative $t \cdot dg^2/dt = - \mu^2 \cdot dg^2/d \mu^2$ i.e.~the usual beta function with an infinitesimal change in the RG scale at any particular $g^2$ value. Note that asymptotic freedom corresponds to $t \cdot dg^2/dt > 0$. In comparison to the approach at fixed $c$ in Section~\ref{gradient}, the flow time $t$ is not held fixed relative to the lattice size $L/a$ in the new method as described in what follows. From a sequence of ensembles with various lattice volumes, fermion masses and lattice spacings, a sequence of limits can be taken to reach the continuum infinitesimal-step beta function in infinite volume in the chiral limit. 

We have previously generated a large set of such ensembles in our study of the particle spectrum of two flavor sextet ${\mathrm {SU(3)}}$ gauge theory. We use staggered fermions with stout link improvement and the Symanzik gauge action in generating the gauge configurations as described in~\cite{Fodor:2012ty}. Our previous lattice studies of the model found a set of massless Goldstone bosons in the chiral limit separated from massive vector, axial vector and baryonic states, with an emergent light scalar, as well as strong evidence that the chiral condensate is non-zero at zero fermion mass \cite{Wong:lat17beta,Fodor:2016pls,Fodor:2012ty}. These p-regime gauge ensembles, already strongly indicative of near-conformal behavior, provide the basis for this beta function computation.

\begin{figure}[thb] % no figure before 1st section
  \centering
 \sidecaption
  \includegraphics[width=5cm,clip]{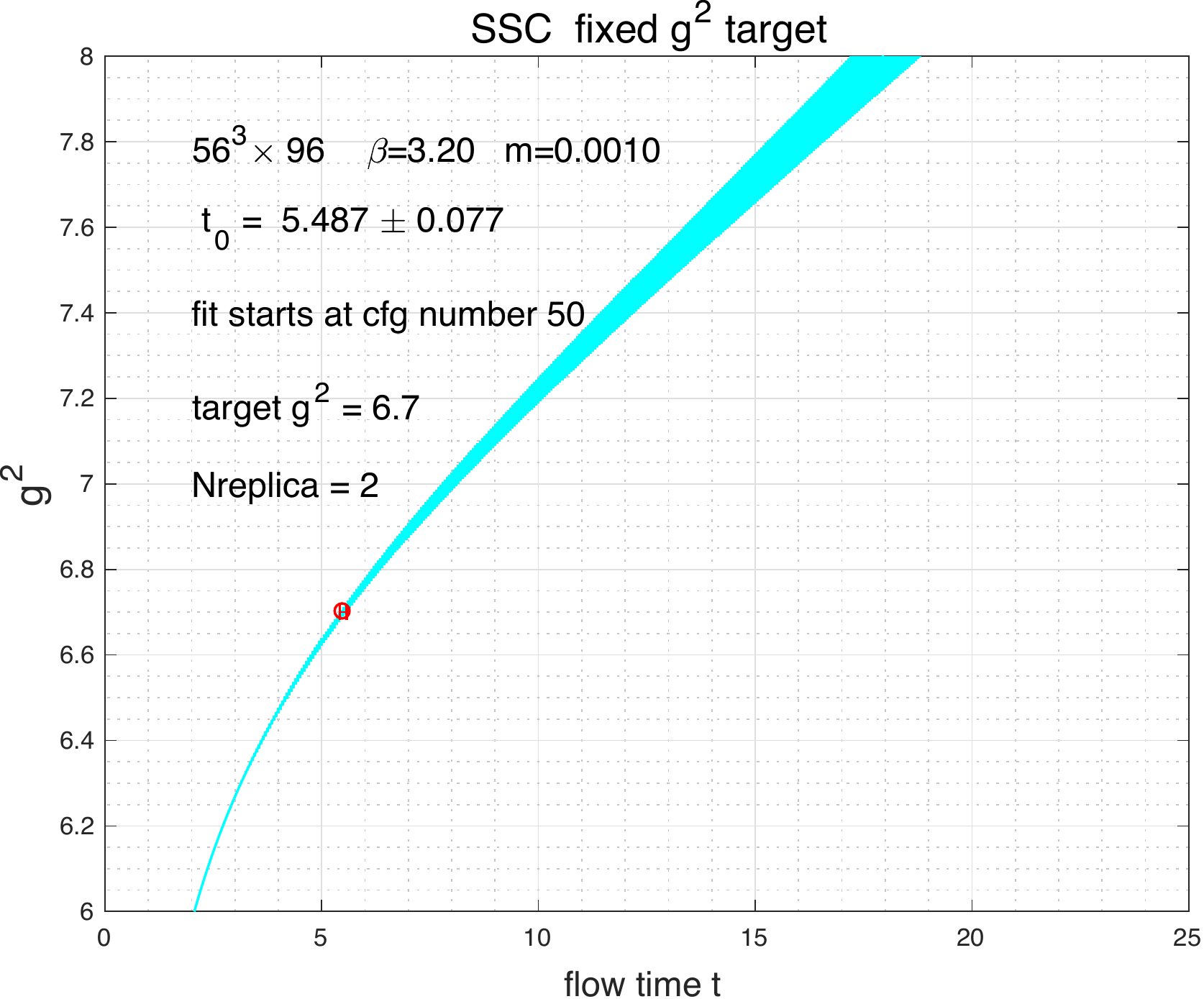}
  \includegraphics[width=5cm,clip]{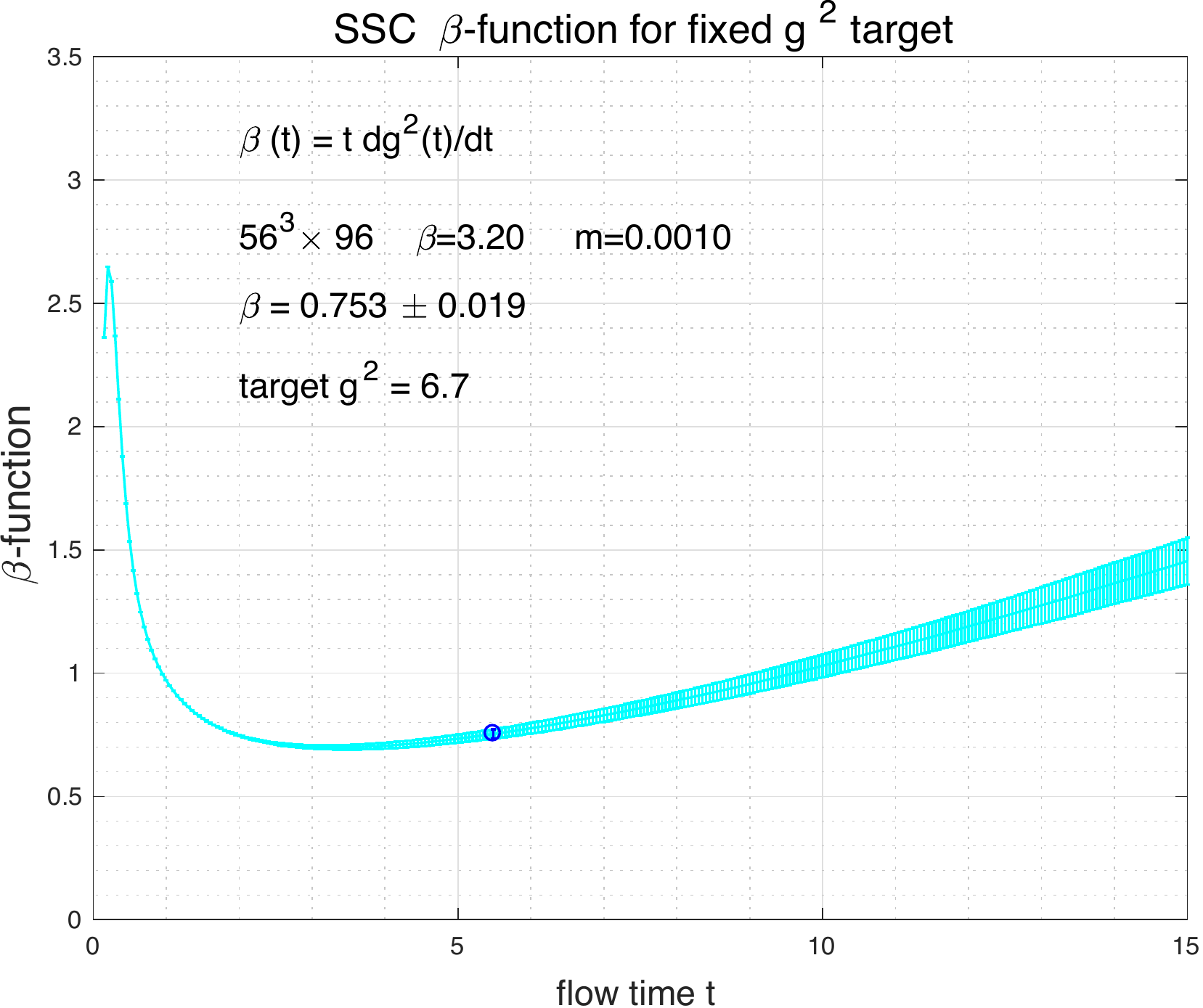}
  \caption{(left) The gradient flow renormalized coupling $g^2$ and (right) its associated beta function on a lattice volume $56^3 \times 96$ at a Goldstone boson mass of $m_{\pi}\cdot a \approx 0.08$.}
  \label{fig-1}% Give a unique label
\end{figure}

In Figure~\ref{fig-1} we show the renormalized coupling $g^2$ and its corresponding derivative $t \cdot dg^2/dt$ for one ensemble, a lattice volume $56^3 \times 96$ at the bare gauge coupling $6/g_0^2 = 3.20$ and fermion mass $ma = 0.001$, corresponding to a Goldstone boson mass $m_{\pi}\cdot a \approx 0.08$. The derivative is approximated by $\{ -F(t+2\ep) + 8 F(t+\ep) - 8 F(t - \ep) + F(t - 2\ep) \}/(12 \ep) = dF/dt + {\cal O}(\ep^4)$. As opposed to step-scaling where the flow time $t$ is set by the choice of $c = \sqrt{8t}/L$, in this method the value of the renormalized coupling $g^2$ is chosen and the flow time where this value is reached is measured. We show the choice $g^2(t_0) = 6.7$, which for this ensemble occurs at $t_0/a^2 = 5.487 \pm 0.077$. (Note that this does not correspond to the choice of $t_0$ set by $t^2 \cdot \langle E \rangle_{t_0} = 0.3$ in the original investigation of~\cite{Luscher:2010iy}.) A larger choice of $g^2$ gives a larger statistical error on $t_0$, however too small a value of $g^2$ gives a beta function distorted by large cutoff effects, as seen on the right of Figure~\ref{fig-1} for $t < 2$. These and other constraints we describe later influence which fixed value of $g^2(t_0)$ we choose to target.

\begin{figure}[thb] % no figure before 1st section
  \centering
 \sidecaption
  \includegraphics[width=5.2cm,clip]{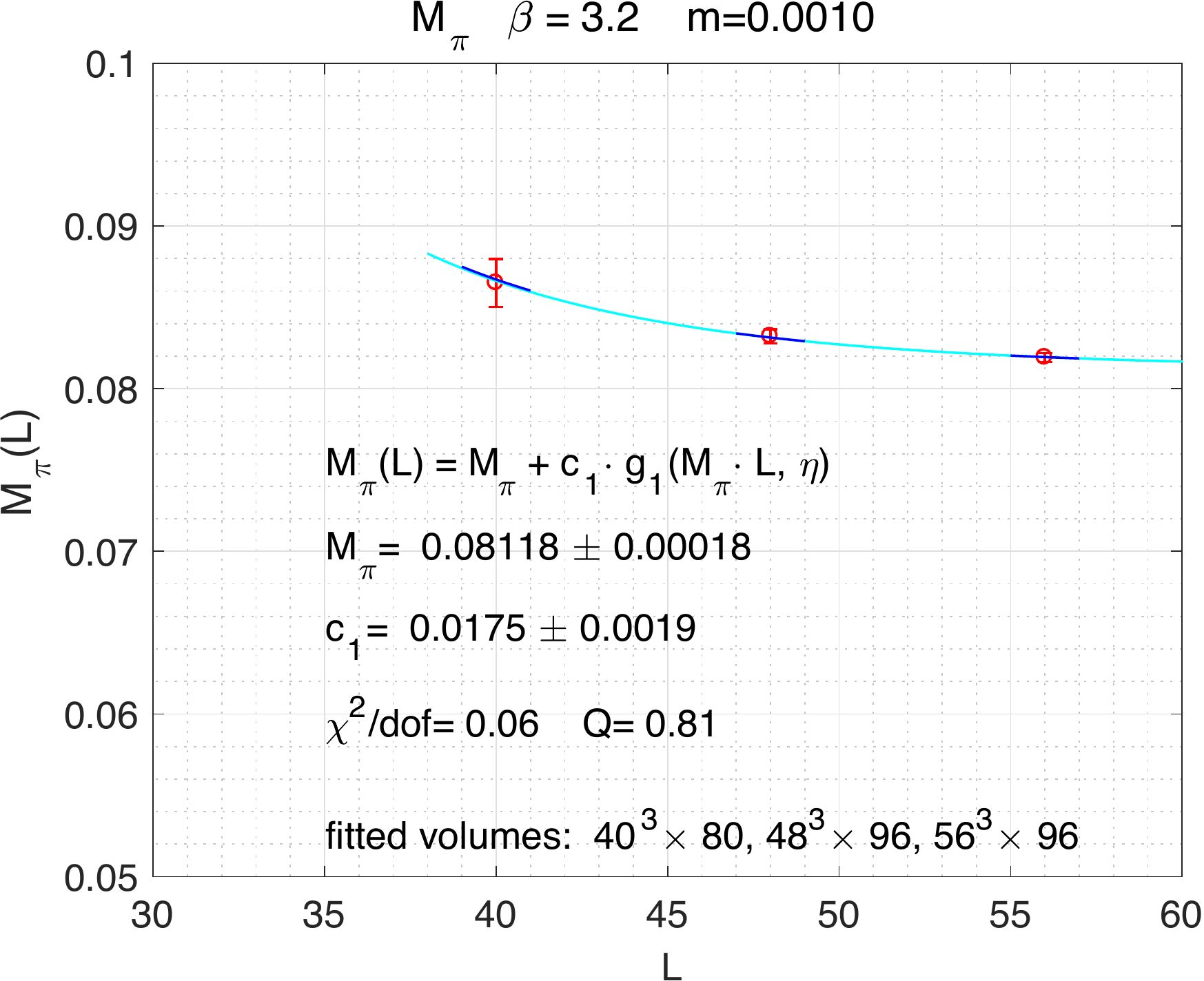}
  \includegraphics[width=5cm,clip]{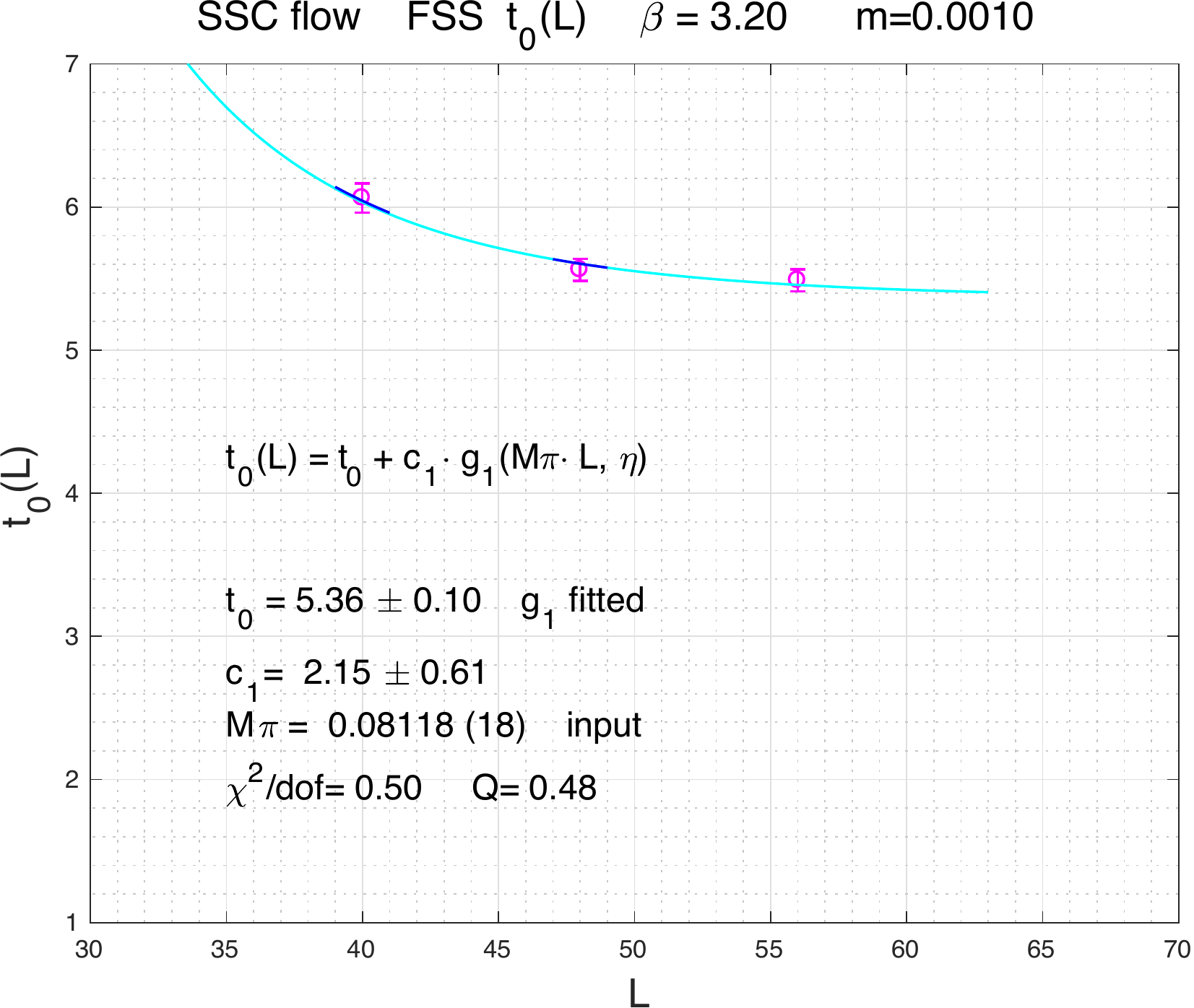}
  \caption{Infinite volume extrapolations of (left) the Goldstone boson mass and (right) the scale $t_0$ at which $g^2(t_0) = 6.7$, at fixed fermion mass and bare coupling.}
  \label{fig-2}% Give a unique label
\end{figure}

\begin{figure}[thb] % no figure before 1st section
  \centering
 \sidecaption
  \includegraphics[width=5cm,clip]{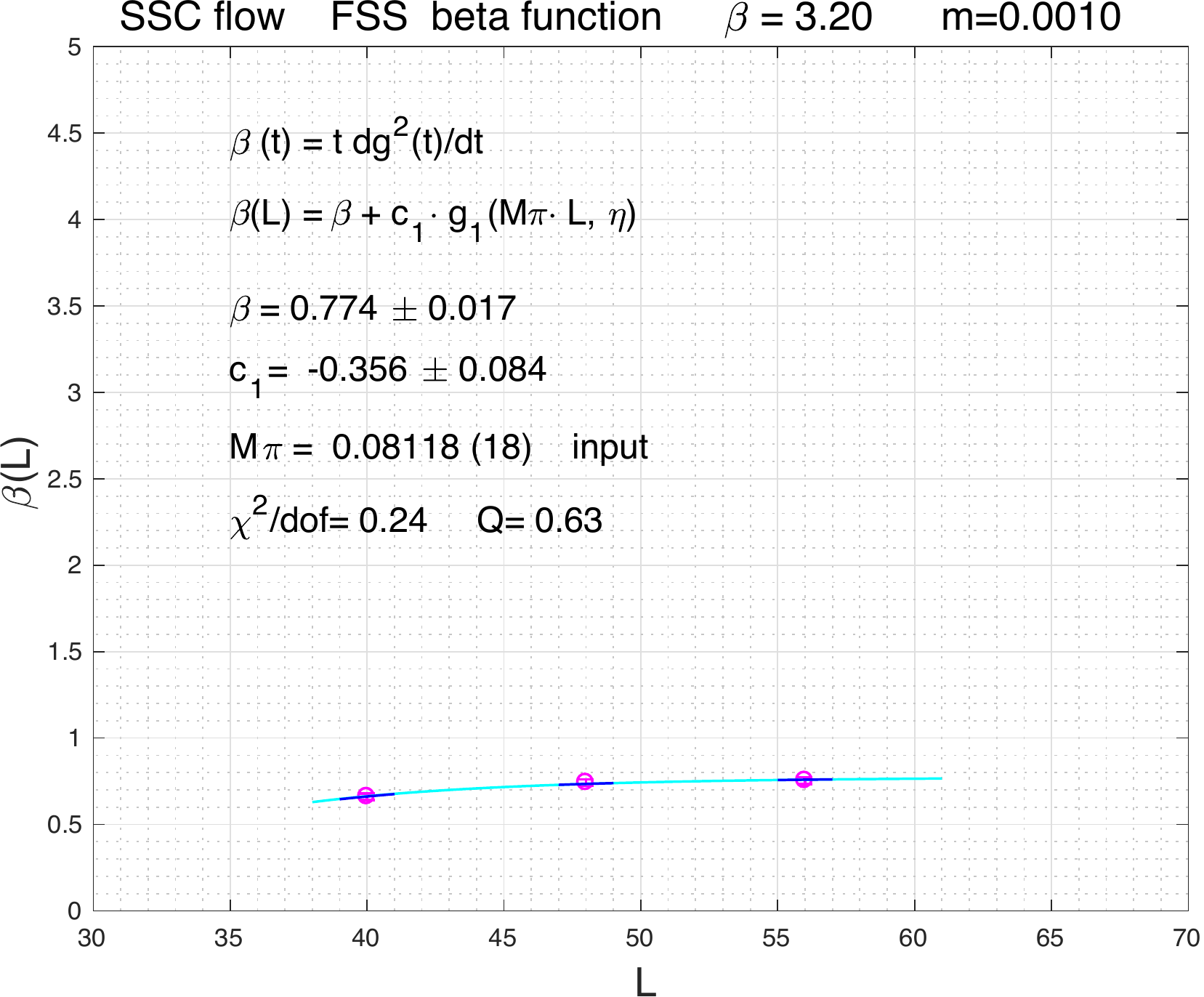}
  \includegraphics[width=5cm,clip]{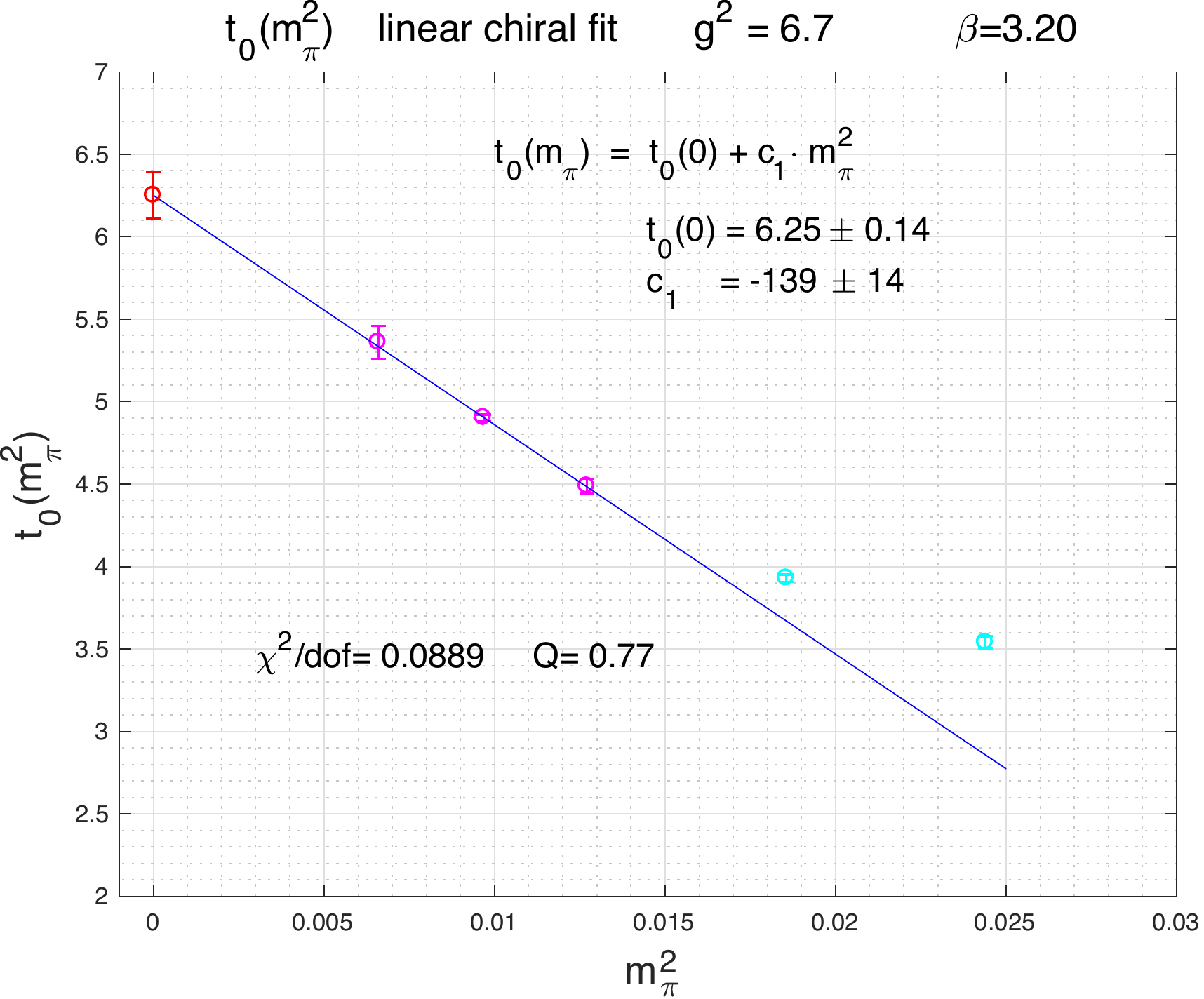}
  \caption{(left) Infinite volume extrapolation of the beta function at the renormalized coupling $g^2(t_0) = 6.7$. (right) Chiral extrapolation of the scale $t_0$ as a function of $M_\pi^2$. The cyan data points are not included in the fit.}
  \label{fig-3}% Give a unique label
\end{figure}

Since the goal is the infinite volume beta function, it is necessary to correct for finite volume dependence. We use an ansatz with an infinite sum $g_1$ of Bessel functions dependent on the aspect ratio $L_t/L_s$ of the lattice volume to account for Goldstone bosons wrapping around the finite volume~\cite{Gasser:1986vb}~e.g. $M_\pi(L) = M_\pi + c_M g_1(M_\pi L)$
where the complicated sum $g_1$ is evaluated numerically. At 1-loop in chiral perturbation theory $c_M = M_\pi^2/(64 \pi^2 F_\pi^2)$, we leave the prefactor $c_M$ of the $g_1$ function as a free parameter to be fitted. In Figures~\ref{fig-2} and~\ref{fig-3} we show examples of such infinite volume extrapolations for the Goldstone boson mass, the scale $t_0$ and the corresponding beta function. These figures are typical: the volume effect is relatively small but visible and is well described by the ansatz. Note that the infinite volume mass $M_\pi$ is first determined by the Goldstone boson volume fit and is then used as one of the inputs for the $t_0$ and beta function volume fits.

\begin{figure}[thb] % no figure before 1st section
  \centering
  \includegraphics[width=6cm,clip]{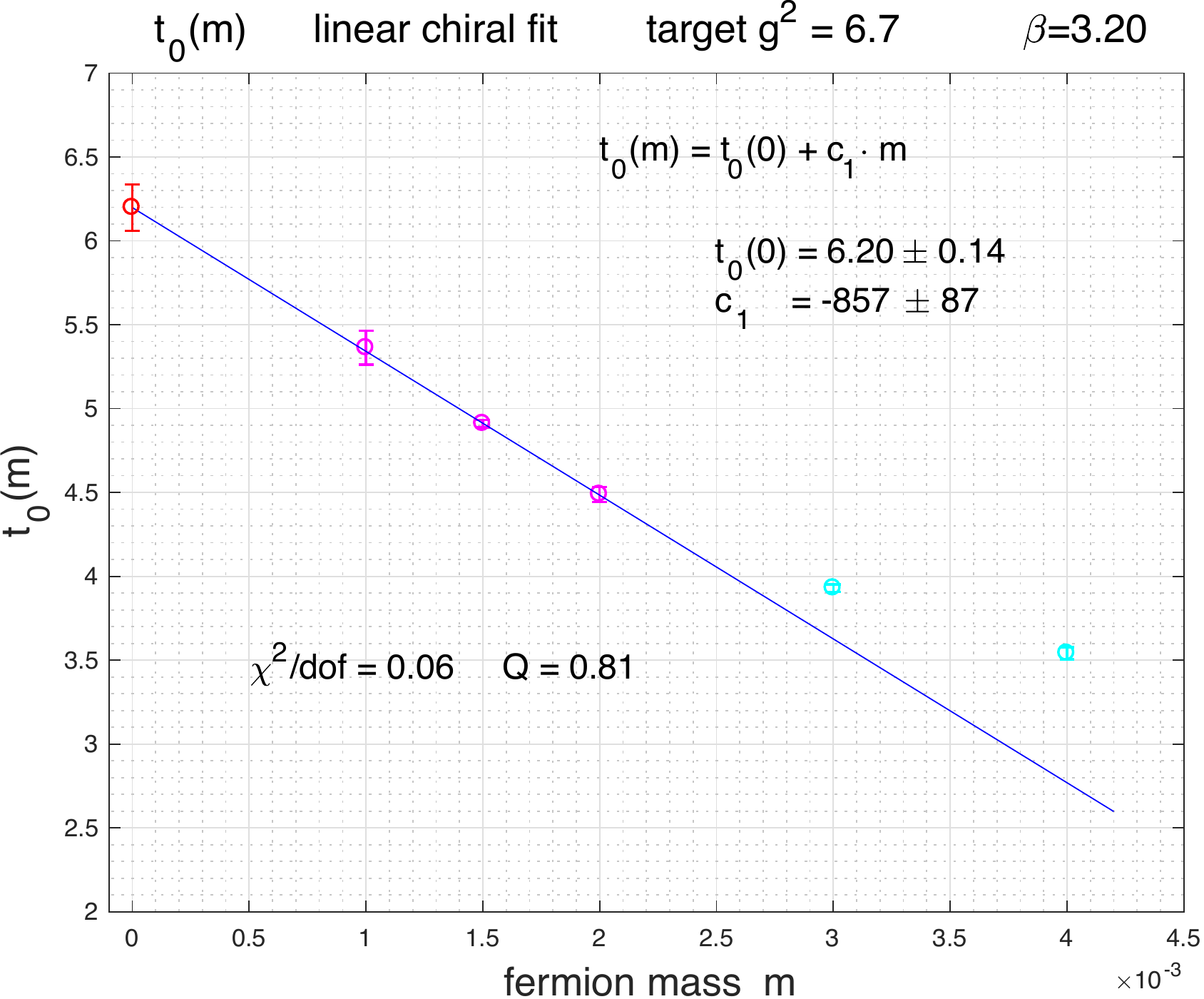}
  \includegraphics[width=6cm,clip]{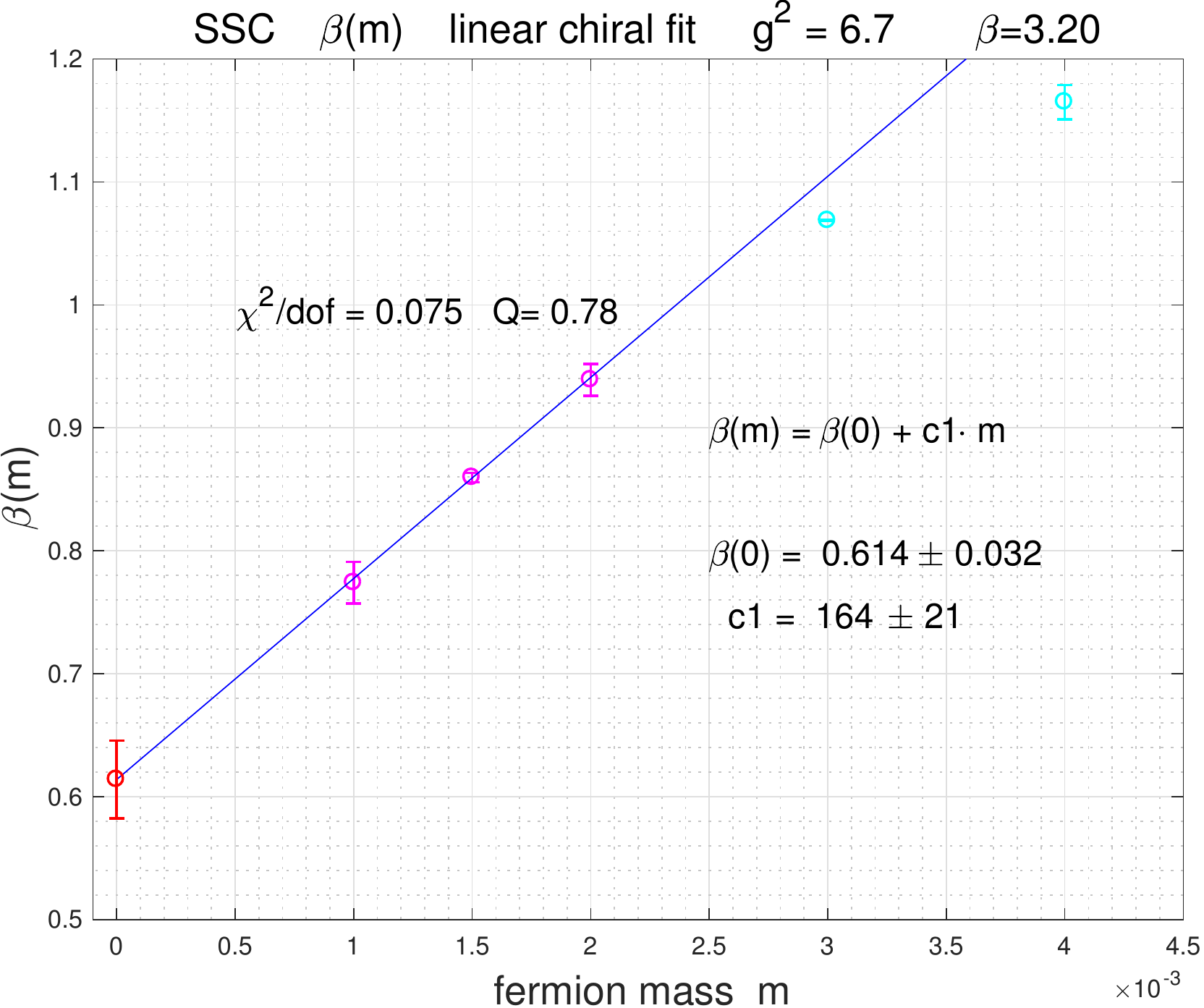}
  \caption{Chiral extrapolations of (left) the scale $t_0$ and (right) the beta function in the fermion mass $m$.}
  \label{fig-4}% Give a unique label
\end{figure}

The next natural step is the extrapolation to zero fermion mass at fixed bare coupling. From~\cite{Bar:2013ora} if the smearing radius $\sqrt{8t}$ is small compared to the Goldstone boson Compton wavelength, a chiral expansion gives  
\be
t_0 = t_{\rm 0,ch} \left( 1 + k_1 \frac{M_\pi^2}{(4 \pi f)^2} + k_2 \frac{M_\pi^4}{(4 \pi f)^4} \log \left( \frac{M_\pi^2}{\mu^2} \right) + k_3 \frac{M_\pi^4}{(4 \pi f)^4} \right)
\label{eq3}
\ee
where $f$ is the Goldstone boson decay constant in the chiral limit. We show in Figure~\ref{fig-3} an example of such a chiral fit of the infinite-volume $t_0$ data. We do not have sufficient data at all lattice spacings for a quadratic fit in $M_\pi^2$ or to fit the chiral logarithm, hence we use a linear fit in $M_\pi^2$ for the data at the lighter masses. At this leading order, linear dependence in $M_\pi^2$ is equivalent to linear dependence in the fermion mass $m$ itself, extrapolating in either variable to the chiral limit should give consistent results. We show in Figure~\ref{fig-4} the results of linear fits in the mass $m$ at the same bare coupling, which are indeed consistent with extrapolating in $M_\pi^2$. The determination of the scale in the chiral limit is $t_0/a^2 = 6.20 \pm 0.14$ at this bare coupling $6/g_0^2 = 3.20$, which corresponds to our coarsest lattice spacing.

\begin{figure}[thb] % no figure before 1st section
  \centering
  \includegraphics[width=6cm,clip]{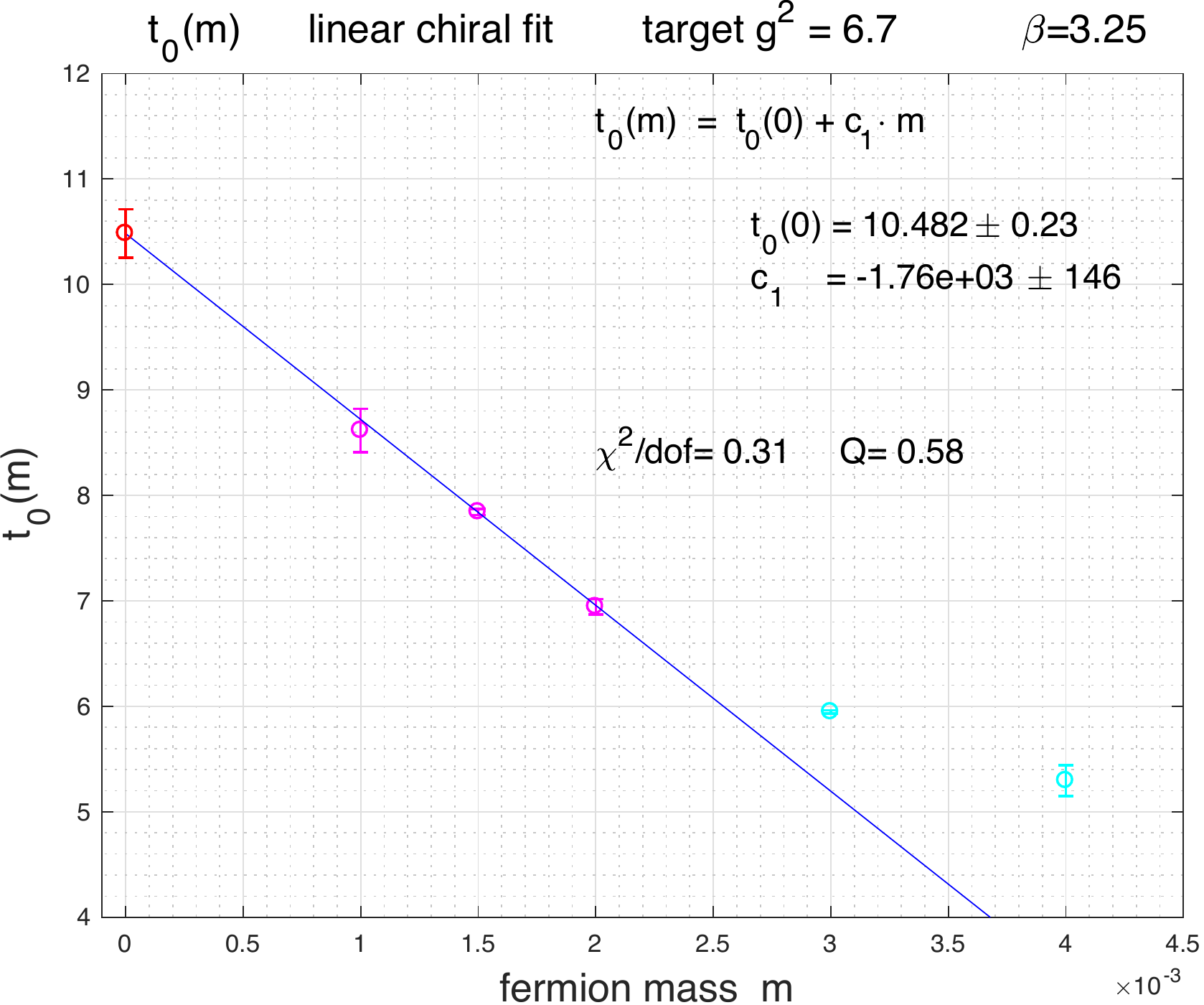}
  \includegraphics[width=6cm,clip]{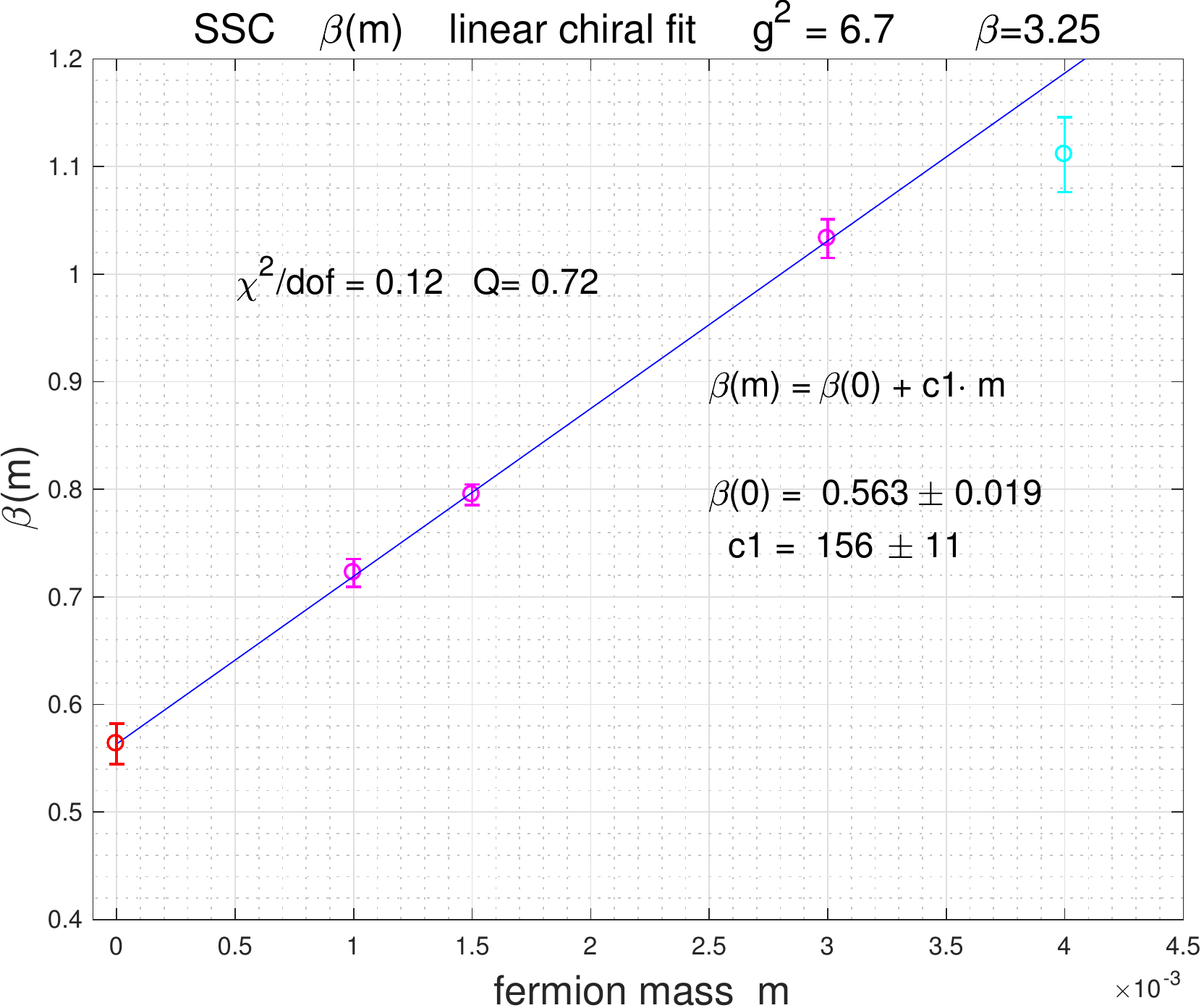}
  \caption{Similar to Figure~\ref{fig-4}, chiral extrapolations at $6/g_0^2 = 3.25$, our intermediate lattice spacing.}
  \label{fig-5}% Give a unique label
\end{figure}

\begin{figure}[thb] % no figure before 1st section
  \centering
  \includegraphics[width=6cm,clip]{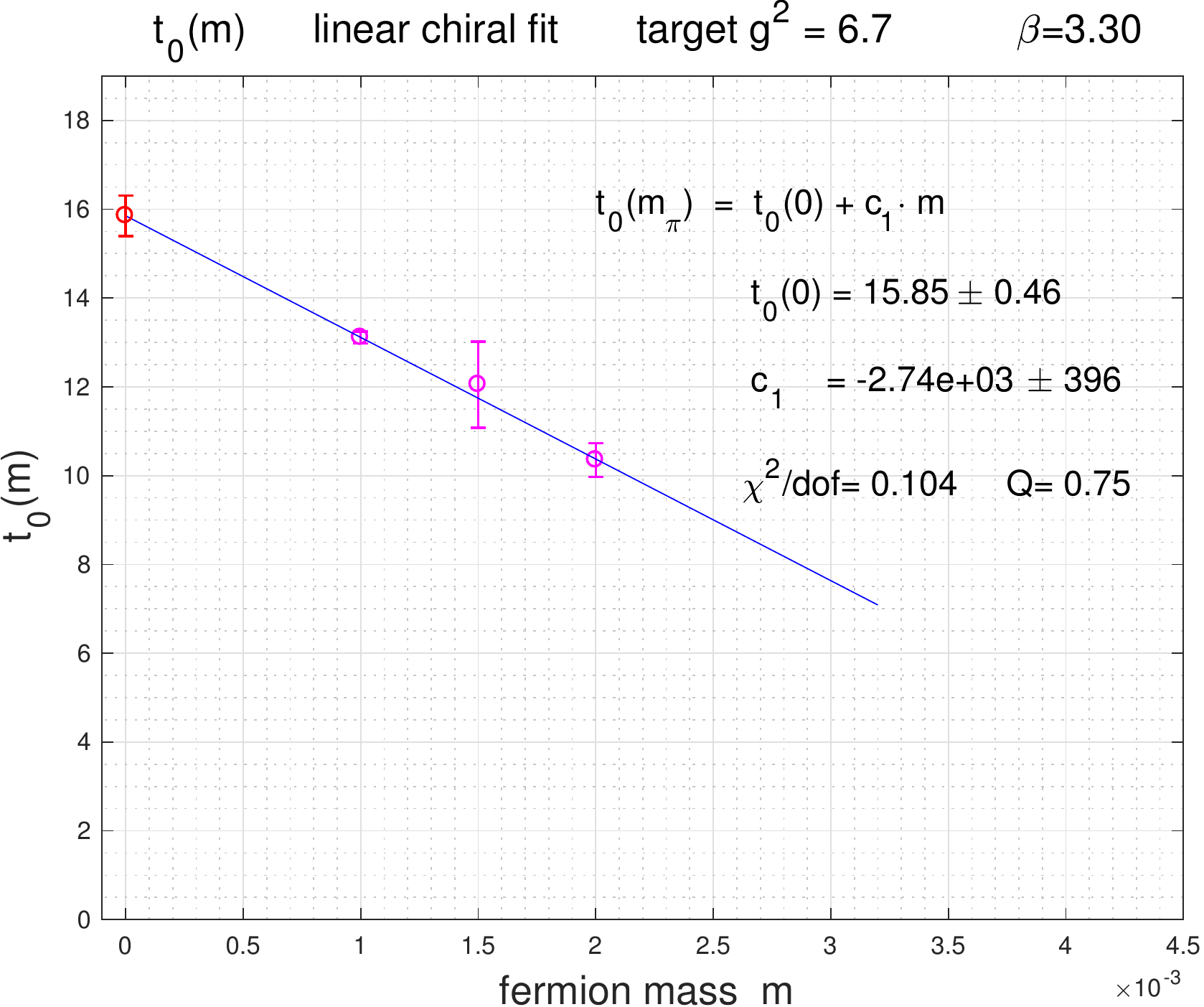}
  \includegraphics[width=6cm,clip]{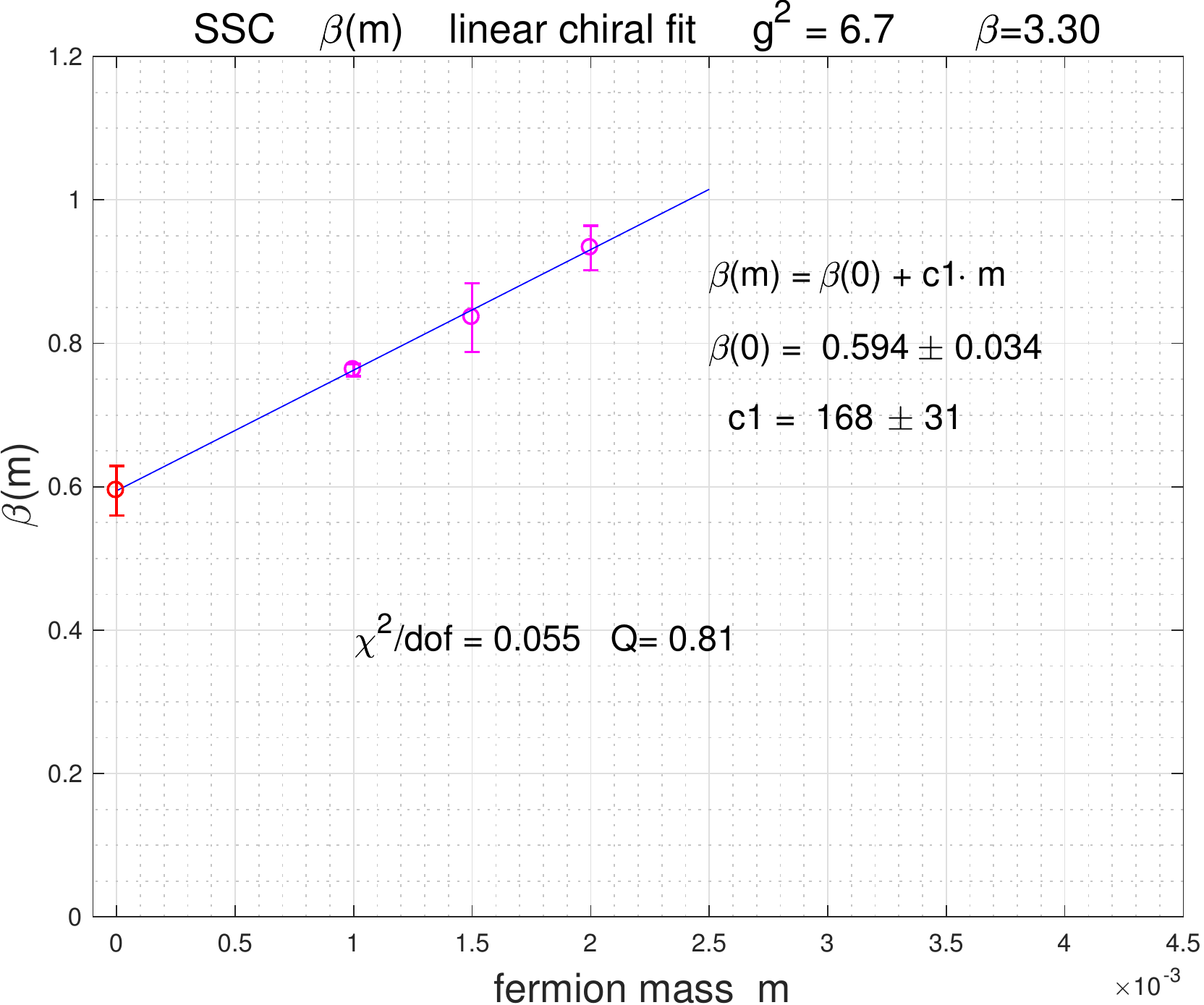}
  \caption{Similar to Figures~\ref{fig-4} and~\ref{fig-5}, chiral extrapolations at $6/g_0^2 = 3.30$, our finest lattice spacing.}
  \label{fig-6}% Give a unique label
\end{figure}

The entire procedure is repeated for two other sets of ensembles: $6/g_0^2 = 3.25$ corresponding to our intermediate lattice spacing, and $6/g_0^2 = 3.30$, our finest lattice spacing. We hold the renormalized coupling $g^2(t_0) = 6.7$ fixed, find the corresponding $t_0/a^2$ and beta function values for a variety of lattice volumes and fermion masses, fit their finite-volume dependence at fixed mass and then extrapolate to the chiral limit. The final step is shown in Figures~\ref{fig-5} and ~\ref{fig-6}. We see that estimates of the chiral limit scale $t_0/a^2$ are $10.48 \pm 0.23$ and $15.85 \pm 0.46$ for the intermediate and fine lattice spacings respectively, giving an overall change of $\approx 1.6$ in lattice spacing from coarsest to finest ensembles. The chiral limit of the beta function shows modest cutoff effects on the order of 10\%, which makes the continuum extrapolation mild. Note that a larger choice of the renormalized coupling to define the scale e.g.~$g^2(t_0) = 8$ would give a larger value of $t_0/a^2$, which might not be possible to accommodate at the finest lattice spacing such that the finite-volume dependence could be removed. On the other hand too small a value of $g^2(t_0)$ would give much larger lattice artifacts, hence the choice $g^2(t_0) = 6.7$ balances these two considerations.

We show the last step, the continuum extrapolation of the beta function, in Figure~\ref{fig-7}. In the chiral limit we expect the leading cutoff effect to be ${\cal O}(a^2)$, hence we fit the data linearly in $a^2/t_0$, with only three data points a more extended fitting form is not possible. Because the fitting variable $t_0$ has its own error, this effect in included in the fit as described in~\cite{0957-0233-18-11-025}, with the $\chi^2$ function being generalized to include the error in both $x$ and $y$ coordinates
\be
\chi^2 = \sum_{k=1}^n \left[ \frac{(X_k - x_k)^2}{\sigma_{x,k}^2} +  \frac{(Y_k - y_k)^2}{\sigma_{y,k}^2} \right],
\label{eq-chi2}
\ee 
where $x_k$ and $y_k$ are the data pairs with their respective errors $\sigma_{x,k}$ and $\sigma_{y,k}$, and $Y_k = c \cdot X_k + d$ is the fitting form with $c$ and $d$ as the parameters to be determined. Using this form, our result for the infinite-volume infinitesimal beta function at $g^2 = 6.7$ is $\beta(g^2) = 0.548 \pm 0.047$. 
Any physical target, like the beta function in this work, requires appropriate orders of the chiral and continuum limits as noted in~\cite{Bernard:2004ab}. An alternative to the approach presented here would take the chiral and continuum limits simultaneously in terms of $\sqrt{t_0} \cdot m$ and $a^2/t_0$, similar to~\cite{Wong:lat17beta}. This method is being investigated for the beta function. 

\section{Comparison and conclusion}\label{conclude}

\begin{figure}[thb] % no figure before 1st section
  \centering
  \includegraphics[width=6cm,clip]{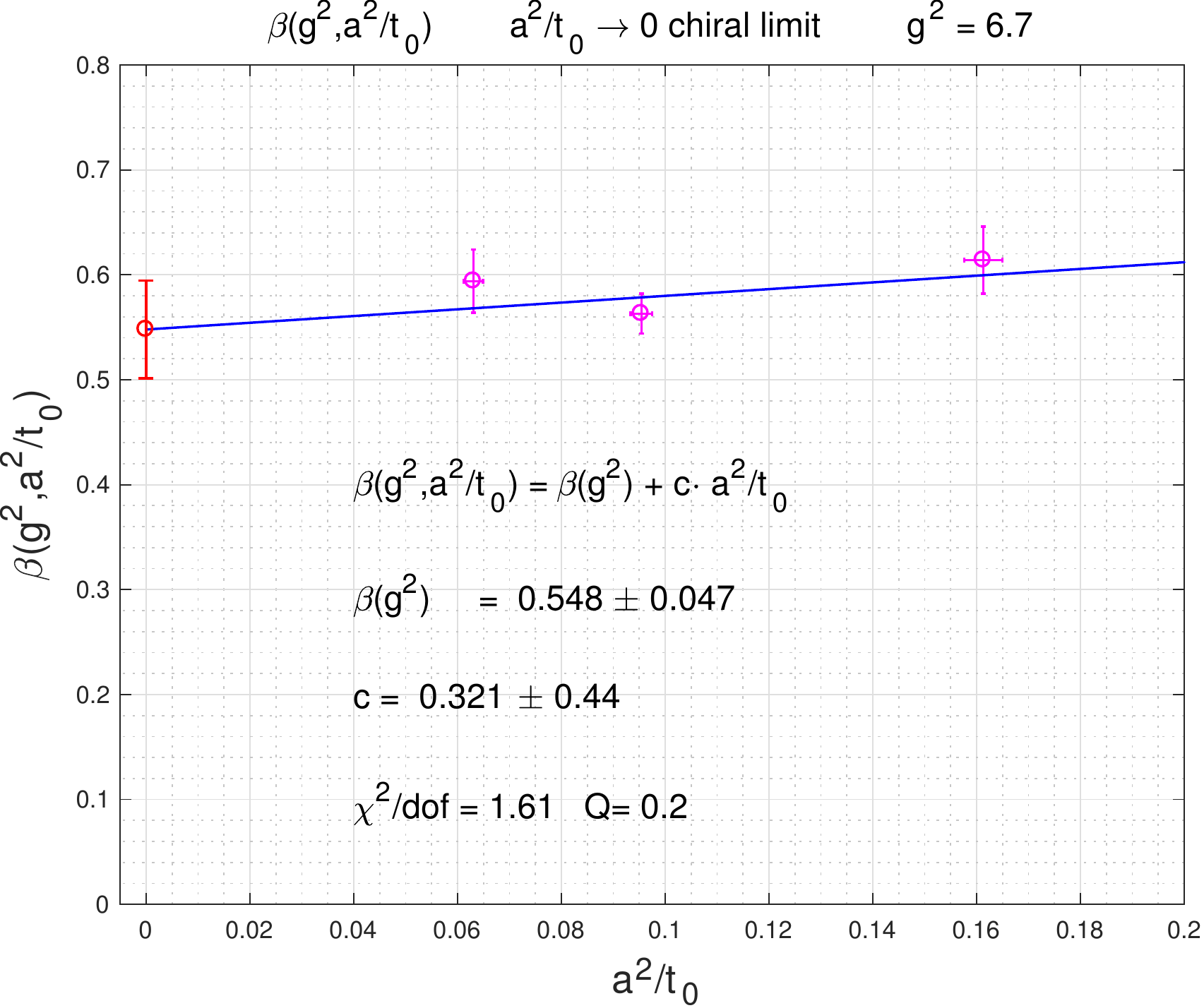}
  \includegraphics[width=7cm,clip]{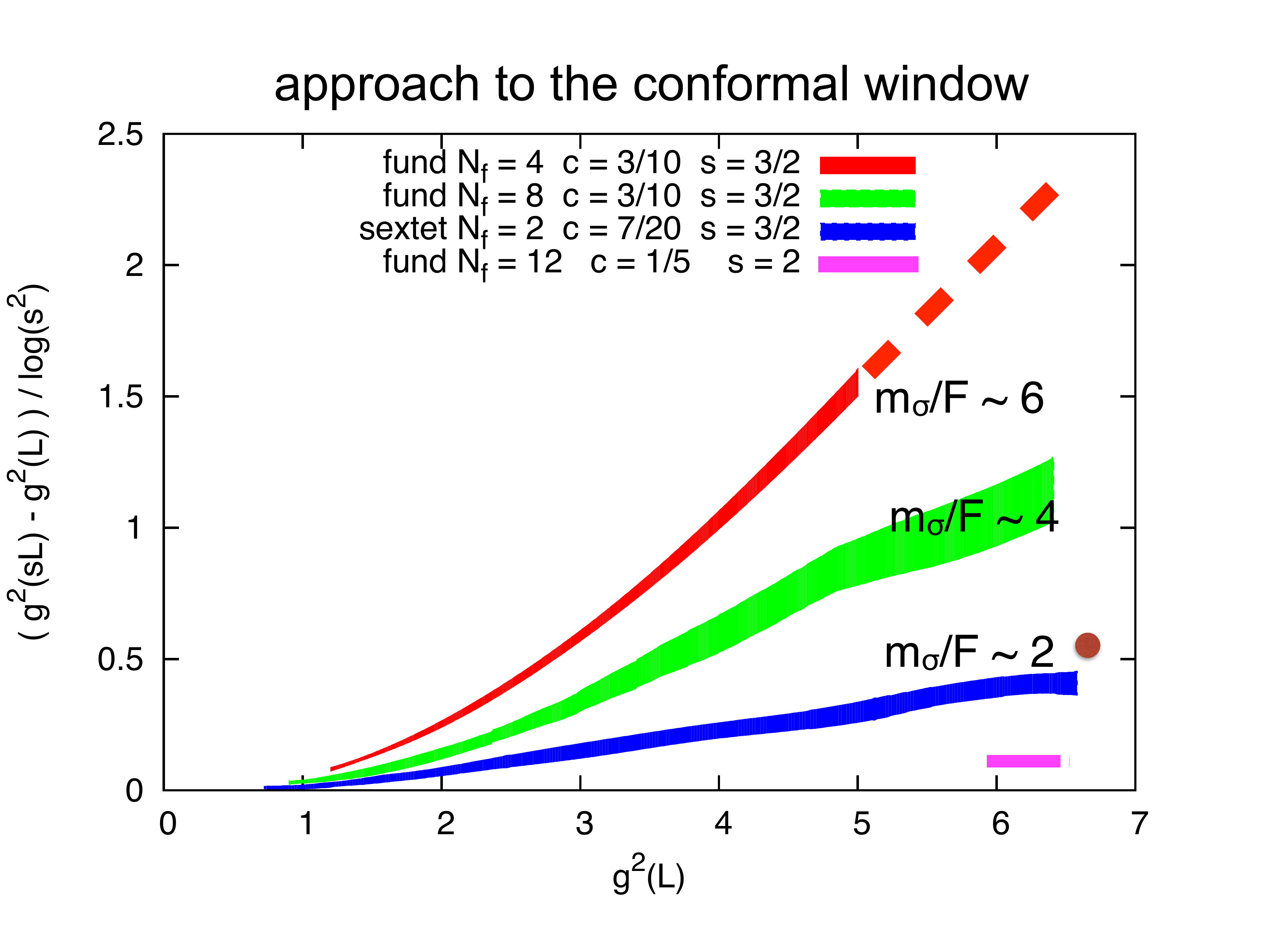}
  \caption{(left) Continuum extrapolation of the beta function at $g^2(t_0) = 6.7$, yielding $\beta = 0.548 \pm 0.047$ as the continuum result. (right) Comparison of this calculation with previous finite volume beta function measurements. In the gradient flow scheme in infinite volume, the 3-loop beta function~\cite{Harlander:2016vzb} has an infrared fixed point at $g^2 \approx 6.8$, in the $\overline{\mathrm {MS}}$ scheme the corresponding 3-loop beta function has a zero at $g^2 \approx 6.3$.}
  \label{fig-7}% Give a unique label
\end{figure}

The infinite volume beta function we determine is in a different scheme than the finite volume beta function measured via step-scaling, which in turn has its own dependence on the choice of $c$, the ratio of flow time to lattice volume. It is still instructive to compare these different results for the sextet model as shown in Figure~\ref{fig-7}, where the finite volume beta function is taken from our own work in~\cite{Fodor:2015zna}. We see that the two calculations are in good agreement -- the beta function is small but non-zero in the range of renormalized couplings which, from our independent studies of the particle spectrum, are strong enough that chiral symmetry is spontaneously broken in the chiral limit. Our recent extended study of the beta function of the twelve-flavor ${\mathrm {SU(3)} }$ model with fundamental representation fermions~\cite{Fodor:2017gtj} shows that at small values of $c$ there is little volume dependence in the method of Section~\ref{gradient}. This may explain the good agreement between our infinite and finite volume beta functions at $g^2 = 6.7$ in the sextet model since the new beta function in some sense might be viewed as the $c \rightarrow 0$ limit.

The finite volume beta function, calculated directly at zero mass, starts in the perturbative regime and moves to stronger coupling as the physical volume grows. If no infrared fixed point (IRFP) is found i.e.~a non-trivial zero of the beta function, one could argue it is simply because strong enough coupling and large enough physical volumes have not yet been reached. However, the gauge ensembles where the finite volume beta function at $g^2 = 6.7$ could be attained are matched by p-regime gauge configurations at the same coupling for the targeted scale but with massless fermions in the infinite volume limit and spontaneous chiral symmetry breaking. This is demonstrated by the particle spectrum and the eigenvalues of the Dirac operator. In this phase the theory has sufficiently strong coupling to generate a p-regime with massive states separated from the massless Goldstone bosons, there is no room left at stronger coupling for the theory to have a conformal spectrum of massless states whose mass deformation would be governed by a universal anomalous dimension. This bridges the gap between the weak and strong coupling regimes and obviates any need to continue exploring even stronger coupling with the finite volume beta function in the hunt for an IRFP.

Our beta function calculations, consistent with one another, contradict other lattice studies of the finite volume beta function for the sextet model~\cite{Shamir:2008pb,Hasenfratz:2015ssa}. We believe this is because of lattice artifacts whose effects were not fully removed in those works. The range of lattice volumes we employ is larger than in either of those studies, which allows us to push further towards the continuum. This is mostly an issue of systematic errors, not a question of underestimated statistical errors, and should be accounted for without any speculation about differing universality classes for different fermion discretizations, contrary to the claims made in~\cite{Hasenfratz:2017mdh}. Our beta function determinations are also consistent with our large-volume non-perturbative study of the particle spectrum, which shows that chiral symmetry is spontaneously broken in the massless fermion limit, with associated Goldstone bosons and a spectrum of massive states~\cite{Wong:lat17beta,Fodor:2016pls,Fodor:2012ty}. This is inconsistent with other studies of the sextet model using Wilson fermion discretization, which interpret the sextet model as being infrared conformal~\cite{Hansen:2017ejh}.

In comparison to ${\rm SU(3)}$ gauge theory with $N_f$ massless fermion flavors in the fundamental representation, the sextet model appears to have near-conformal behavior, with a lighter composite scalar than in the $N_f=4$ and 8 theories. Our first investigations of the anomalous mass dimension, measured via the Dirac operator eigenvalues, indicates that it could be sufficiently large to be phenomenologically viable~\cite{Fodor:2016hke}. If this first sign holds, and is combined with the other properties of the sextet model, the theory continues to be a relevant and interesting candidate for explicit realization of the composite Higgs paradigm. However the entangled dynamics of the light scalar and the light Goldstone pion with need for a generalized framework in chiral perturbation theory remains an unsolved problem. This is under active investigation as addressed in~\cite{Kuti:lat17dila} with potential implications for the beta function analysis presented here.

\section*{Acknowledgments}
We acknowledge support by the DOE under grant DE-SC0009919, by the NSF under grants 1318220 and 1620845, by OTKA under the grant OTKA-NF-104034, and by the Deutsche
Forschungsgemeinschaft grant SFB-TR 55. Computational resources were provided by the DOE INCITE program on the ALCF BG/Q platform, by USQCD at Fermilab, by the University of Wuppertal, by Juelich Supercomputing Center on Juqueen and by the Institute for Theoretical Physics, Eotvos University. We are grateful to Szabolcs Borsanyi for his code development for the BG/Q platform. We are also grateful to Sandor Katz and Kalman Szabo for their CUDA code development.

\bibliography{holland2017}

%%%%%%%%%%%%%%%%%%%%%%%%%%%%%%%%%%%%%%%%%%%%%%%%%%%%%%%%%%%%%%%%%%%%%%%%%%%%%
\end{document}